\documentclass[doublecol]{epl2} 

\usepackage{latexsym}

\usepackage{graphicx}
\usepackage{amssymb}
\usepackage{epstopdf}
\usepackage{float}
\title{Wave Resistance for Capillary Gravity Waves: Finite Size Effects}
\shorttitle{Wave Resistance for Capillary Gravity Waves: Finite Size Effects}
\author{M. Benzaquen\inst{1} \and F. Chevy\inst{2} \and E. Rapha\"el\inst{1}}
\shortauthor{M. Benzaquen \etal}

\institute{                    
  \inst{1} Laboratoire PCT - UMR  Gulliver CNRS 7083, ESPCI, 10 rue Vauquelin, 75005 Paris, France\\
  \inst{2} Laboratoire Kastler Brossel - ENS, Universit\'e Paris 6, CNRS, 24 rue Lhomond, 75005 Paris, France
}


\abstract{
We study theoretically the capillary-gravity waves created at the water-air interface by an 
external surface pressure distribution symmetrical about a point and
 moving at constant velocity along a linear trajectory. 
Within the framework of linear wave theory and assuming the fluid 
to be inviscid, we calculate the wave resistance experienced by the
perturbation as a function of its size (compared to the capillary length). In particular,
we analyze how the amplitude of the jump occurring at the minimum phase speed
$c_{{\rm min}}=(4 g \gamma /\rho)^{1/4}$ depends on the size of the pressure distribution
($\rho$ is the liquid density, $\gamma$ is the water-air surface tension, 
and $g$ is the acceleration due to gravity).
We also show how for pressure distributions broader than a few capillary lengths, the result obtained
by Havelock for the wave resistance in the particular case of pure gravity waves ({\it{i.e.}}, $\gamma = 0$) is 
progressively recovered.}

\begin{document}

\maketitle

\section{Introduction}

Water waves are both captivating and of great practical significance \cite{Lighthill, Lamb, Johnson}. 
They have thus attracted the attention of scientists and engineers for many decades \cite{Darrigol}. 
Water waves can, for instance, be generated by the wind blowing over the ocean, by a moving ship on a calm lake, 
or simply by throwing a stone into a pond.  Their propagation at the surface of water is driven by a balance between the liquid
inertia and its tendency, under the action of gravity or of surface tension (or a combination of both in the
case of {{\it capillary-gravity waves}}), to return to a state of
stable equilibrium \cite{LandauLifshitz}. The dispersive nature of water  waves is responsible for
the complicated wave pattern generated at the free surface
of a still liquid by a moving disturbance such as
a partially immersed object (e.g. a boat or an insect)
or an external surface pressure source \cite{Acheson}. The propagating
waves generated by the moving disturbance continuously
remove energy to infinity. Consequently, the disturbance
will experience a drag, $R$, called the {\it{wave resistance}}. In the case of 
ships (for which surface tension is negligible), this drag is known to be a major source of resistance \cite{Milgram}
and has been analyzed in detail by Havelock \cite{Havelock}.
The case of objects that are small compared
to the capillary length has been considered
only recently \cite{Elie:96, Elie:99,  Keller, Chevy,
 Bacri, Steinberg1,Alexei, Closa} and has attracted strong interest in the context of
insect locomotion on water surfaces \cite{Deny, Bush, Voise}.
For such objects, one has to take into account both gravity  and surface tension.
In the case of a point-like surface pressure distribution, this leads to a discontinuity
of the wave resistance at a critical velocity given by the minimum of the wave velocity
$c_{\rm min}=(4g\rho/ \gamma)^{1/4}$ \cite{Elie:96}.
For clean water at room temperature, one has  $c_{\rm min} \approx 0.23 \mathrm{\, m \,s^{-1}}$.
When the velocity $V$ of the surface pressure distribution is smaller than 
 $c_{\rm min}$,   no  steady waves are generated
and the wave resistance vanishes. Emission of steady waves becomes possible
only when $V > c_{\rm min}$, leading to the onset of a finite wave drag \cite{rem1}.
The wave resistance discontinuity 
at the critical velocity $c_{\rm min}$ has been experimentally investigated by several groups \cite{Bacri, Steinberg1}. 

It is important to notice that while the analysis of \cite{Elie:96} was mainly concerned with
a point-like surface pressure distribution \cite{overlooked},
real perturbations (like insects) have finite sizes.  The aim of the present paper
is thus to analyze in detail the role played by the finite size of the surface pressure distribution
 for the wave resistance \cite{RefASteinberg}. We will see in particular that for $b$ much larger 
than $\kappa^{-1}$, the influence of surface tension is negligible. In that case,
the results of Havelock \cite{Havelock} for pure gravity waves are recovered.

\section{Formulation}

Consider an incompressible, inviscid, infinitely deep liquid whose free surface is unlimited. We take the $xy$-plane as the equilibrium surface of the fluid and the $z$-axis along the upward direction perpendicular to the equilibrium surface.
We study the wave motion created by an external 
surface pressure distribution that moves with speed $V$ in the negative $x$-direction. In the frame of that moving
disturbance, physical quantities are stationary: the pressure distribution is given by $P(x,y)$ and 
the displacement $\zeta$  of the free surface from its equilibrium position  is of the form $\zeta(x,y)$.

In order to calculate the wave resistance experienced by the disturbance, we use a method first introduced by Havelock
 \cite{Havelock}. According to this author, we may imagine a rigid cover fitting the surface everywhere. The assigned pressure system $P(x,y)$ is applied to the liquid by means of this cover; hence the wave resistance is simply the total resolved pressure in the $x$ direction. This leads to 
\begin{eqnarray}
R=\int dx\,dy\,P(x,y)\partial_x\zeta(x,y).
\end{eqnarray}

For the sake of simplicity, let us restrict ourselves to the case of a pressure system symmetrical around the origin so that $P(x,y)=g(r)$ with $r=\left(x^2+y^2\right)^{1/2}$). The Fourier transform $\hat P(k_x,k_y)$ is then a function only of $k$ and can be written as $\hat P (k_x,k_y)=G(k)$, where
\begin{eqnarray}
G(k)&=&\int_0^\infty dr\,rg(r)\int_0^{2\pi} d\theta \,e^{-ikr\cos\theta}\\
&=&2\pi\int_0^\infty dr\,rg(r)J_0(kr).
\label{G}
\end{eqnarray}

$J_0$ denotes the Bessel function of the first kind of zeroth order. It has been shown by  Rapha\"el and De Gennes in \cite{Elie:96} that in such a case the wave resistance $R$ reduces to 
\begin{eqnarray}
R= \int_0^\chi \frac{d\theta\cos\theta}{\pi\gamma} \frac{\{k_{2}(\theta) G[k_2(\theta)]\}^2+\{k_1(\theta)G[k_1(\theta)]\}^2}{k_2(\theta)-k_1(\theta)} \label{Elie1}
\end{eqnarray}
where 
\begin{eqnarray}
k_1(\theta)&=&\kappa\left(\frac V{c_{{\rm min}}}\right)^2\{ \cos^2\theta-(\cos^4\theta-\cos^4\chi)^{1/2} \},\nonumber\\
k_2(\theta)&=&\kappa\left(\frac V{c_{{\rm min}}}\right)^2\{ \cos^2\theta+(\cos^4\theta-\cos^4\chi)^{1/2} \},
\end{eqnarray}
where for $V\geq c_{{\rm min}}$, $\chi$ is  defined by $\cos\chi=c_{{\rm min}}/V$.
Equation (\ref{Elie1}) is an important result as it predicts how the wave resistance varies
as a function of the velocity for any pressure distribution.

\section{Finite size effects on the wave resistance}

Equation (\ref{Elie1}) was studied in detail in  \cite{Elie:96} in the 
 particular case of a point-like pressure distribution  $g(r)=p\,\delta(r)$.
We here consider the case of a pressure distribution of finite size, $b$.
For instance, we can assume the pressure distribution to be Gaussian

\begin{eqnarray}
g(r)=\frac p{2\pi b^2}{\,\exp{ \left(-\frac{r^2}{2b^2} \right)}},\label{Gaussian}
\end{eqnarray}

Equation (\ref{G}) then becomes
\begin{eqnarray}
G(k)=p\exp\left({-\frac{b^2k^2}{2}}\right).\label{GaussTF}
\end{eqnarray}

\begin{figure}[h!]
\begin{center}
\includegraphics[width= 0.85 \columnwidth]{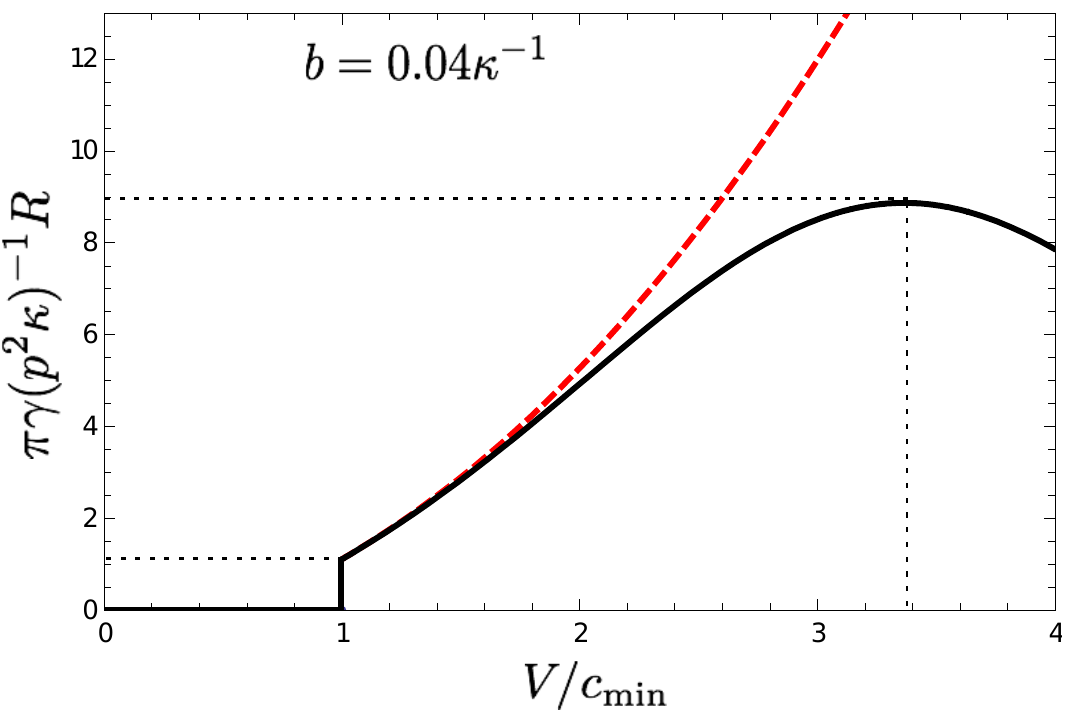}
\end{center}
\caption{(Color online) Plot of the wave resistance $R$  in units
of $p^2 \kappa (\pi \gamma)^{-1} $, as a function of $V/c_{{\rm min}}$. The red (dashed) curve
 corresponds to a point-like pressure source $g(r)=p\,\delta(r)$, the black one (solid line) corresponds to a Gaussian pressure  field (\ref{Gaussian}) of size $b=0.04\, \kappa^{-1}$.} \label{res1}
\end{figure}

\begin{figure}[h!]
\begin{center}
\includegraphics[width= 0.88 \columnwidth]{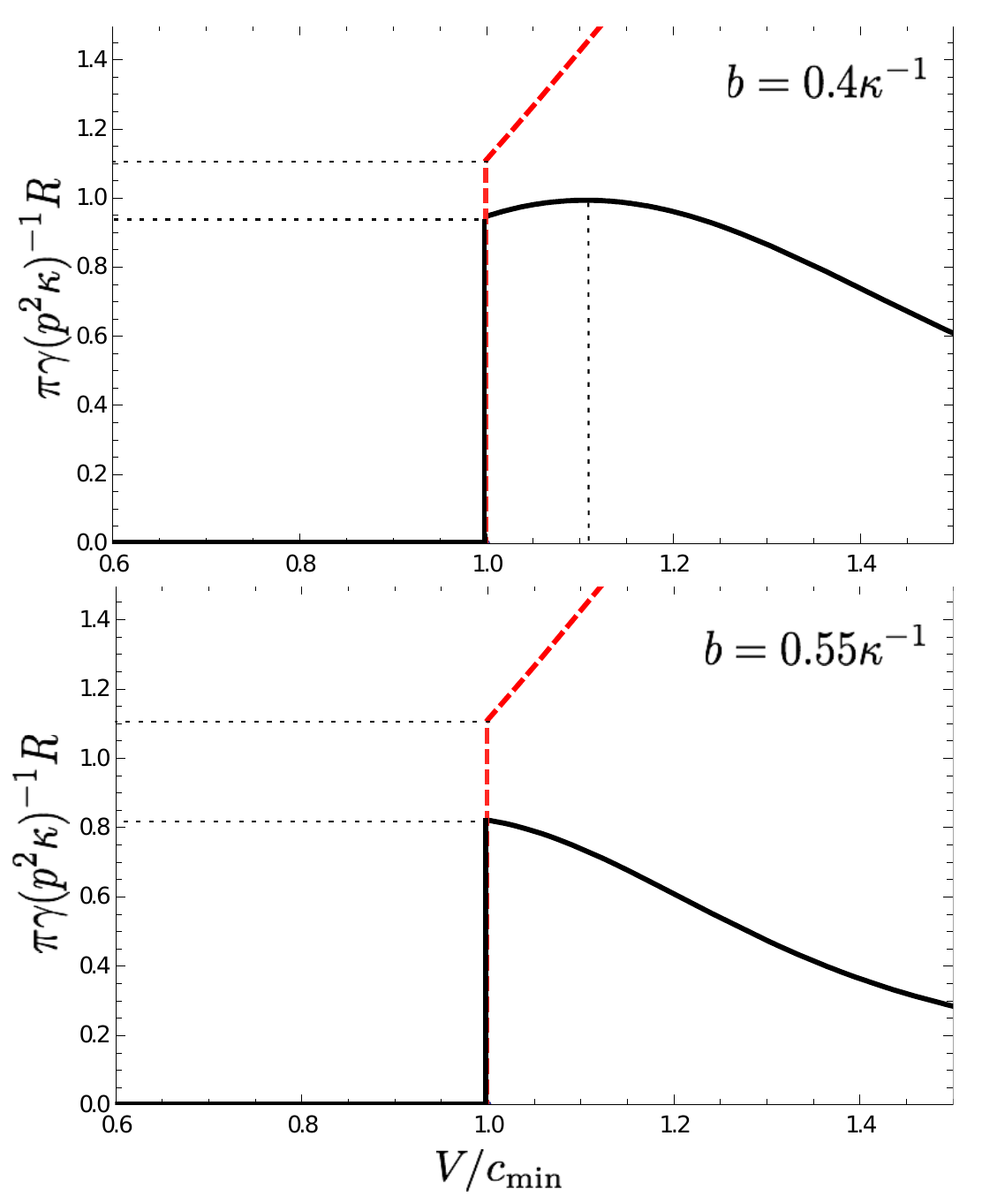}
\end{center}
\caption{(Color online) Plot of the wave resistance $R$  in units
of $p^2 \kappa (\pi \gamma)^{-1} $, as a function of $V/c_{{\rm min}}$. The red (dashed) curve
 corresponds to a point-like pressure field $g(r)=p\,\delta(r)$, the black one (solid line) corresponds to a gaussian pressure  field (\ref{Gaussian}) of size $b=0.4\, \kappa^{-1}$ (upper graph) and $b=0.55\, \kappa^{-1}$ (lower graph).} \label{res1gr}
\end{figure}

By inserting the Fourier transform of the pressure field $G(k)$ in equation (\ref{Elie1}), we obtain
 the wave resistance as a function of $V/c_{\rm min}$ (see Figs.\ref{res1} and \ref{res1gr}).
\smallskip

The red (dashed) in curve in Fig. \ref{res1} corresponds to a point-like pressure distribution $g(r)=p\,\delta(r)$, the black one (solid line) corresponds to a gaussian pressure  field (\ref{Gaussian}) of size $b=0.04\, \kappa^{-1}$. We clearly observe, for both curves, the discontinuity (or jump) at $V=c_{\rm min}$ that we discussed earlier. In Fig.\ref{res1}, the black curve (solid line) presents a maximum at $V_{{\rm max}}\sim\sqrt{\gamma/(\rho b)}$ which is the first consequence of the finite size effects. This separates the behavior of the wave resistance in two regimes: below $V_{{\rm max}}$, $R$ increases with the disturbance velocity whereas for $V>V_{{\rm max}}$, $R$ decreases with $V$.

\smallskip

\begin{figure}[h!]
\begin{center}
\includegraphics[width= 0.88 \columnwidth]{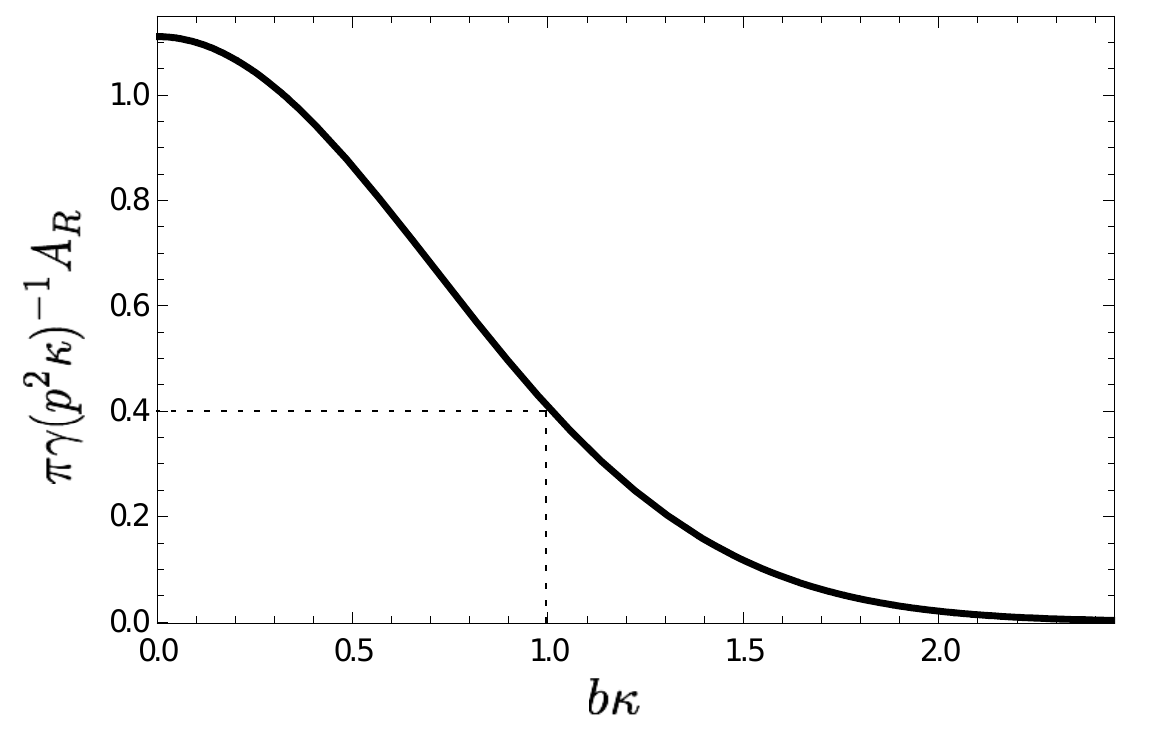}
\end{center}
\caption{Plot of the amplitude of the jump $A_R$ of the wave resistance  in units
of $p^2 \kappa (\pi \gamma)^{-1} $, as a function of $b\kappa$.} \label{res2}
\end{figure}

\begin{figure}[h!]
\begin{center}
\includegraphics[width= 0.88 \columnwidth]{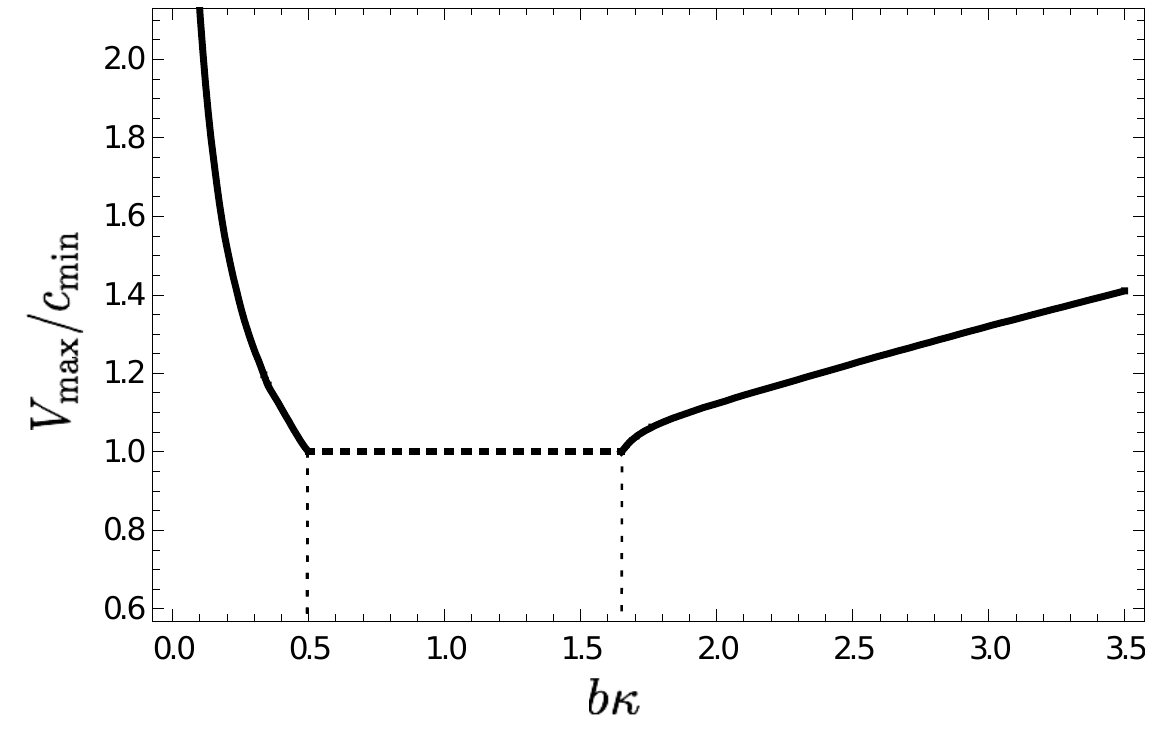}
\end{center}
\caption{Plot of $V_{{\rm max}}$ corresponding to the maximum of wave resistance  in units
of $V/c_{{\rm min}}$, as a function of $b\kappa$. The black dotted  line (horizontal) corresponds to a situation in which the maximum value is reached  at $V_{{\rm max}}=c_{{\rm min}}$. For $b\kappa\ll 1$, $V_{{\rm max}}$ scales as $\sqrt{\gamma/(\rho b)}$, whether for $b\kappa\gg 1$, $V_{{\rm max}}$ scales as $\sqrt{gb}$.} \label{absc}
\end{figure}

\begin{figure}[h!]
\begin{center}
\includegraphics[width= 0.88 \columnwidth]{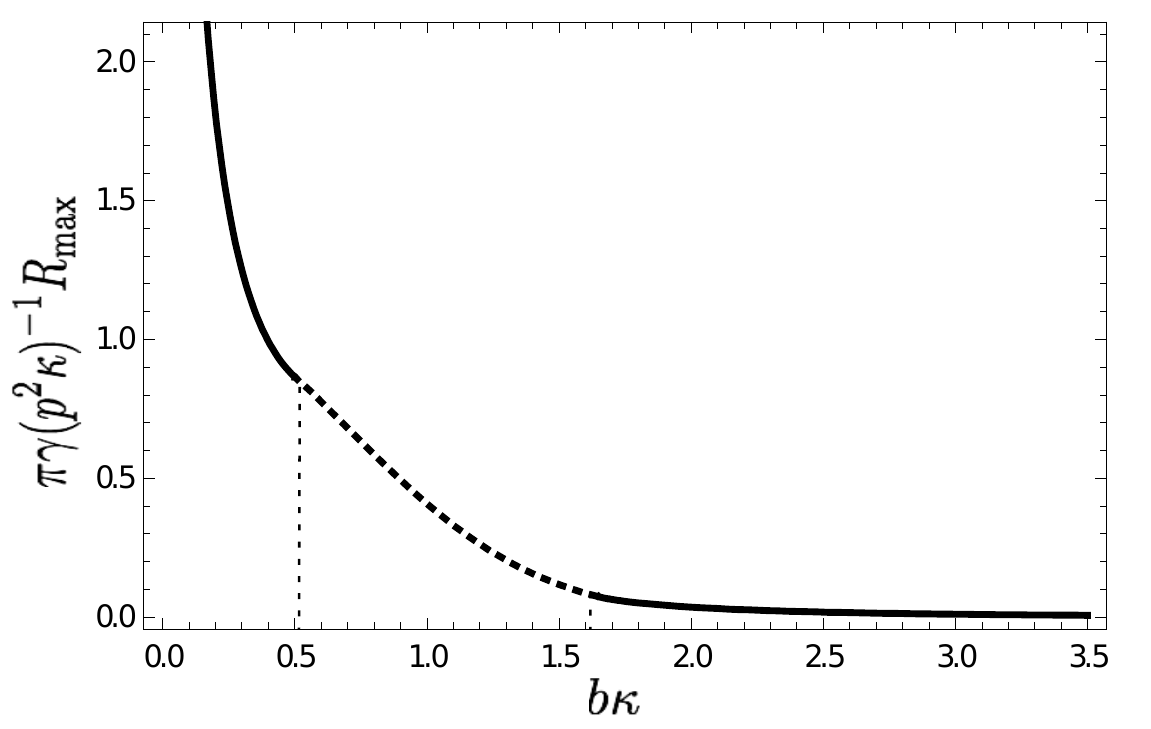}
\end{center}
\caption{Plot of the maximum of wave resistance $R_{{\rm max}}$ in units
of $p^2 \kappa (\pi \gamma)^{-1} $, as a function of $b\kappa$. The dotted line corresponds to a situation in which the maximum value is reached  at $V_{{\rm max}}=c_{{\rm min}}$.} \label{ord}
\end{figure}

In Fig.\ref{res1}, the disturbance typical size $b$  is much smaller than the capillary length $\kappa^{-1}$. When $b$ becomes of the same order of magnitude than $\kappa^{-1}$ (see Fig.\ref{res1gr}), the amplitude of the jump, denoted by $A_R$ in the following, decreases as shown in Fig.\ref{res2}. When the typical size $b$ gets close enough to the capillary length $\kappa^{-1}$, the maximum value of the wave resistance is obtained for  $V_{{\rm max}}=c_{{\rm min}}$. Such a situation is depicted in the lower graph  in Fig.\ref{res1gr}. Figures \ref{absc} and  \ref{ord} give quantitative information on this situation: we show that in the particular case of a Gaussian pressure field it occurs for $0.5\,\kappa^{-1}\lesssim b \lesssim 1.65\,\kappa^{-1}$. 
Note that the above interval is only valid for the Gaussian pressure field (\ref{Gaussian}).
If, for instance,  one uses a Lorentzian pressure source ${p}({2\pi})^{-1}{b}{(b^2 + r^2)^{-3/2}}$ instead
of (\ref{Gaussian}), the above
interval becomes $0.3\,\kappa^{-1}\lesssim b \lesssim 2.2\,\kappa^{-1}$. 
\smallskip

From equation (\ref{Elie1}) on can get the jump amplitude at $V=c_{{\rm min}}$:
\begin{eqnarray}
A_R=\frac{\kappa}{2\sqrt 2}\frac{G(\kappa)^2}{\gamma}=\left( \frac{p^2}{\pi \gamma}\kappa \right)\frac{\pi}{2\sqrt 2}\left( \frac{G(\kappa)}{p} \right)^2,
\label{AR1}
\end{eqnarray}
Equation (\ref{AR1}) is an important result as it gives how the jump of the wave resistance at the critical velocity
$c_{{\rm min}}$ varies with the size of the pressure distribution.
In the particular case of the Gaussian pressure field given in equation (\ref{Gaussian}), the
jump amplitude becomes
\begin{eqnarray}
A_R=\left( \frac{p^2}{\pi \gamma}\kappa \right)\frac{\pi}{2\sqrt 2}\,\exp\left({-b^2k^2}\right).
\end{eqnarray}
Note that in the limit $b \to 0$, the result of \cite{Elie:96} $A_R=\pi/(2\sqrt2)$ for 
of a point-like pressure distribution
is recovered. The amplitude of the jump $A_R$ of the wave resistance  in units
of $p^2 \kappa (\pi \gamma)^{-1} $, is depicted in Fig.\ref{res2} as a function of $b\kappa$. We notably observe that $A_R$ is significantly suppressed when the typical size $b$ becomes greater than $2.5\,\kappa^{-1}$. In such a case, 
the jump in the wave resistance might be difficult to observe experimentally
 (see also  Fig.\ref{ARMAX} below). \smallskip
 
One may also wonder how the maximum in the wave resistance scales with $b$. The abscissa ($V_{{\rm max}}$) and ordinate ($R_{{\rm max}}$) of the maximum of wave resistance as a function of $b\kappa$
 are depicted in Fig.\ref{absc} and Fig.\ref{ord}.

\smallskip

In both figures the dotted line corresponds to the situation in which  the maximum value is reached  at $V_{{\rm max}}=c_{{\rm min}}$  (see lower graph in Fig.\ref{res1gr}).

\smallskip

In the limit $b\kappa\gg 1$, the asymptotic behavior  $V_{{\rm max}} \simeq \sqrt{gb}$  
can be simply recovered by substituting $b^{-1}$ to $k$ in the pure
gravity wave dispersion relation $\omega(k)/k= \sqrt{g/k}$. 
In the opposite limit $b\kappa\ll 1$, the asymptotic behavior  
$V_{{\rm max}} \simeq \sqrt{\gamma/(\rho b)}$  
can analogously be obtained by substituting $b^{-1}$ to $k$ in the pure
capillary wave dispersion relation  $\omega(k)/k= \sqrt{\gamma k/\rho}$ \cite{VmaxDisp}.

\smallskip

We observe in Fig.\ref{ord} that $R_{{\rm max}}$ (and hence the whole wave resistance)
strongly decreases with the size $b$ of the pressure source. Although this might be 
surprising at first sight, one has to notice that we have kept constant the magnitude 
$p = \int dx\,dy\,P(x,y)$ of the pressure source while varying $b$.

\begin{figure}[h!]
\begin{center}
\includegraphics[width= 0.88 \columnwidth]{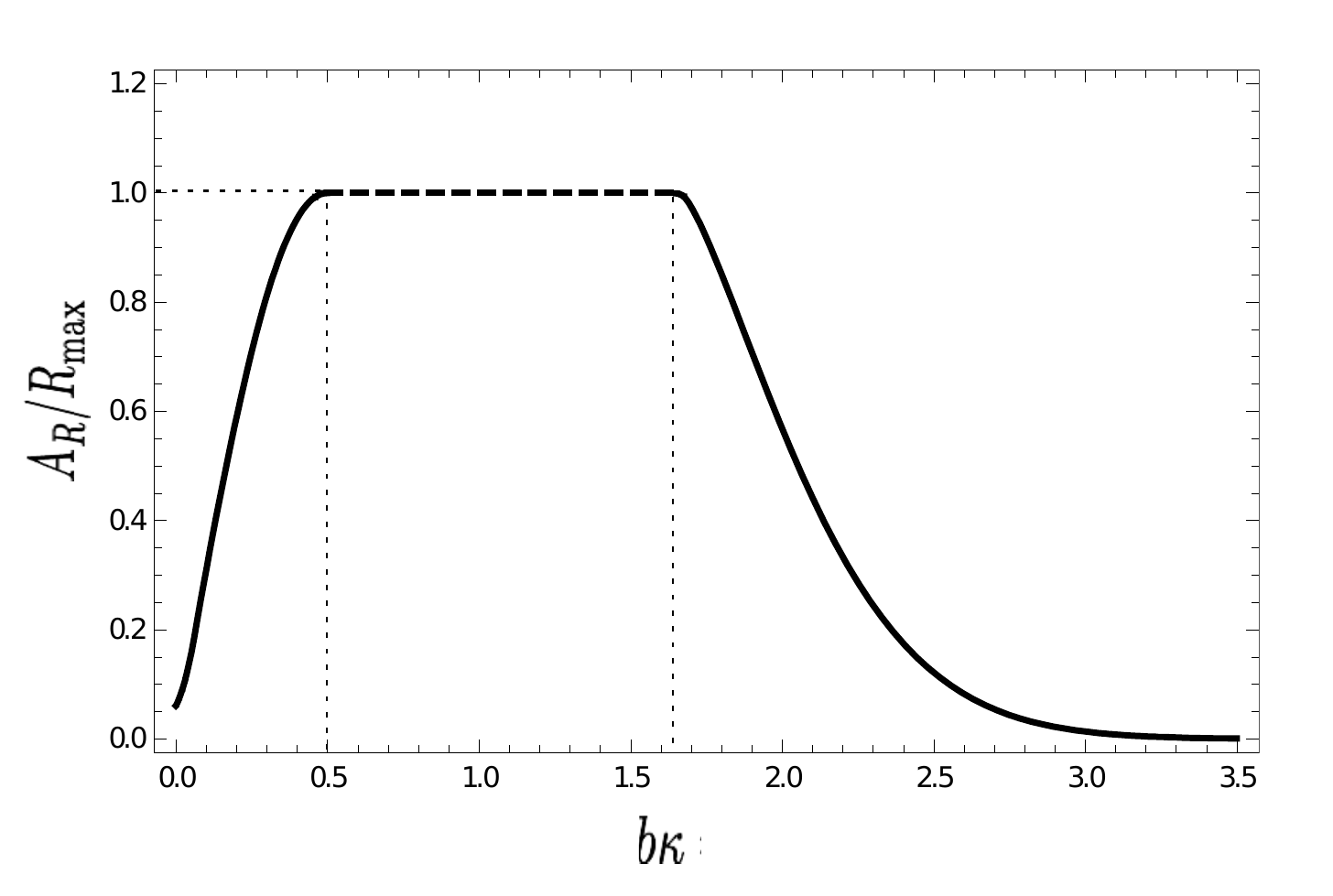}
\end{center}
\caption{Plot of the quotient  $A_R/R_{{\rm max}}$ as a function of $b\kappa$. The dotted line corresponds to a situation in which the maximum value is reached  at $V_{{\rm max}}=c_{{\rm min}}$.} \label{ARMAX}
\end{figure}

\smallskip

In order to discuss the general shape of the graph of the wave resistance, and whether or not the jump in the wave resistance can be easily detected, we plot the ratio between the amplitude $A_R$ and the maximum of wave resistance $R_{{\rm max}}$ (see Fig.\ref{ARMAX}). One can see how the ratio $A_R/R_{{\rm max}}$ starts at zero (for $\kappa^{-1}\rightarrow 0$, $A_R$ is finite and $R_{{\rm max}}\rightarrow \infty$)  and increases until reaching the constant value 1 in the range $0.5\,\kappa^{-1}\lesssim b \lesssim 1.65\,\kappa^{-1}$ discussed earlier  (see Fig.\ref{res1gr}). 
For larger values of $b$, the ratio $A_R/R_{{\rm max}}$ is essentially suppressed.

\section{Pure gravity waves}

As we have seen above,  $V_{{\rm max}}$ scales as $\sqrt{gb}$ for $b\kappa\gg 1$. 
One shall thus wonder if, more generally, our results for the wave resistance 
are for $b\gg \kappa^{-1}$ well approximated by the result of Havelock \cite{Havelock} for pure gravity waves ($\gamma=0$).

\begin{figure}[h!]
\begin{center}
\includegraphics[width= 0.85 \columnwidth]{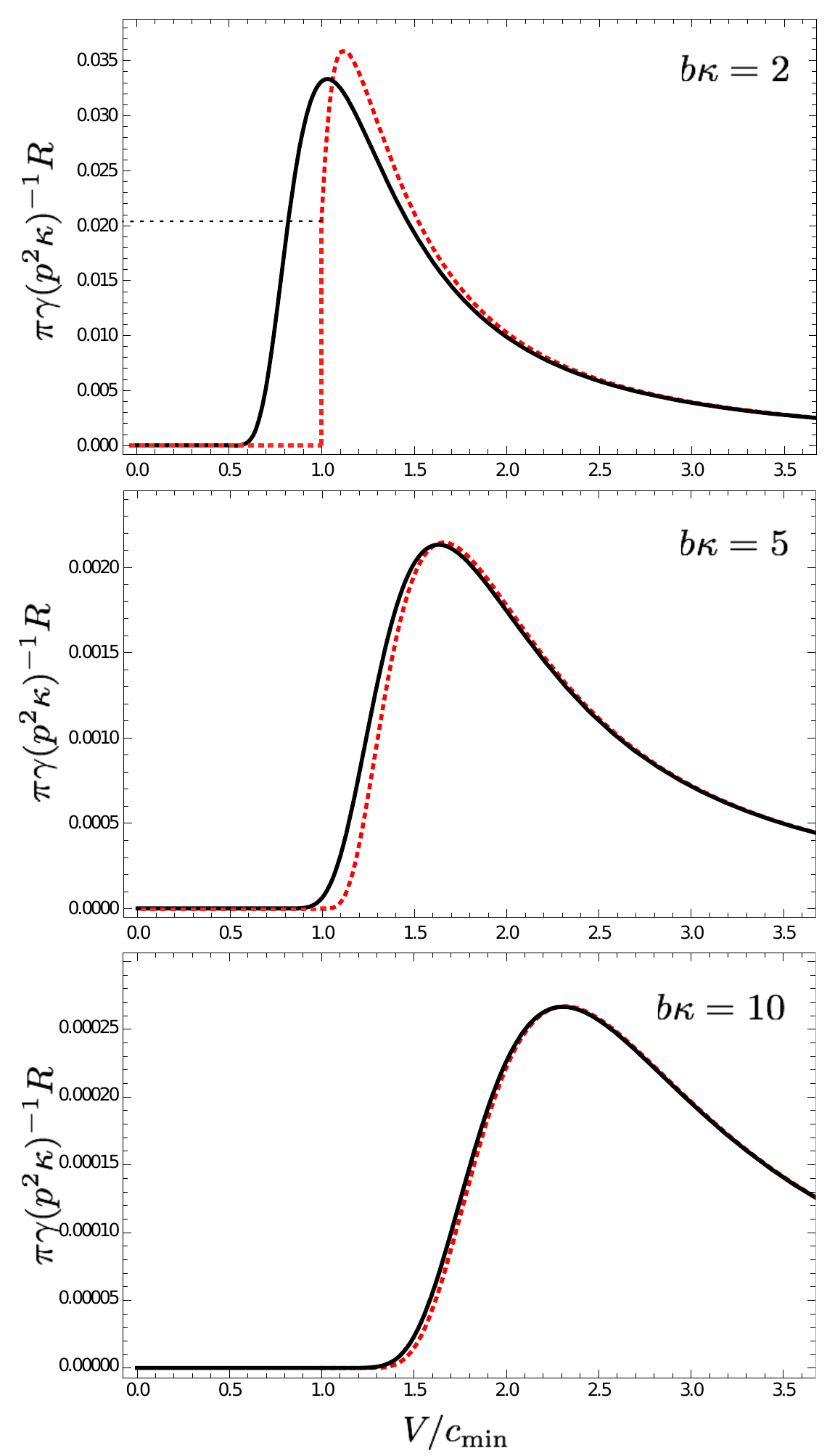}
\end{center}
\caption{(Color online) Plot of the wave resistance $R$  in units
of $p^2 \kappa (\pi \gamma)^{-1} $, as a function of $V/c_{{\rm min}}$. The black curve
(solid line) corresponds to the formula for pure gravity waves (\ref{pure}), the red one (dashed) corresponds to the initial formula for capillary gravity waves (\ref{Elie1}). On the top we have  $b=2\, \kappa^{-1}$, in the middle we have $b=5\,\kappa^{-1}$,  while at the bottom we have $b=10\,\kappa^{-1}$.} \label{respure}
\end{figure}

If $\gamma=0$, the wave resistance (\ref{Elie1}) reduces to
\begin{eqnarray}
R=\frac{g^2}{\rho V^6}\frac{1}{\pi} \int_0^{\frac \pi 2} \frac{d\theta}{\cos^5\theta}\,\left({G\left( \frac{g}{V^2\cos^2{\theta}} \right)}\right)^2,
\end{eqnarray}
\smallskip

which, for the Gaussian pressure field (\ref{Gaussian}), yields
\begin{eqnarray}
R&=&\frac{p^2}{\pi \rho gd^3}\left(\frac{\sqrt{gb}}{V}\right)^6\times\nonumber\\
&&\int_0^{\frac \pi 2} \frac{d\theta}{\cos^5\theta}\,\exp\left({\displaystyle -\frac{1}{\cos^4\theta}\left( \frac{\sqrt{gb}}{V} \right)^4}\right).\label{pure}
\end{eqnarray}

Using equation (\ref{pure}), one can plot  the wave resistance for pure gravity waves as a function of  $V/c_{{\rm min}}$
for different values of $b$
(see  the black curves (solid line) in Fig.{\ref{respure}}). Note that $V_{{\rm max}}$ is now exactly given by $\sqrt{g b}$.
For velocities  smaller than $0.6\sqrt{g b}$, the wave resistance is practically unnoticeable \cite{approx}  . 

We have also plotted in Fig.{\ref{respure}} the wave resistance for capillary-gravity waves (red (dashed) curves).

 One can see how both curves come closer to each other as $b$ is increased. For $b$ much larger than $\kappa^{-1}$, one 
does not see much differences between the two curves, meaning that in this limit the problem is 
essentially ruled by pure gravity waves theory.

\bigskip

\section{Concluding Remarks}

We have shown theoretically how the finite size of an external
axisymmetric pressure source significantly modify the wave resistance, and in particular 
its singular behavior at the minimum phase speed
$c_{{\rm min}}$, compared to the case of a point-like disturbance \cite{nonsym}.
Our study also provides a quantitative description of the crossover between capillary-gravity 
and purely gravity wave resistance described in previous works.

\acknowledgments

We would like to thank A. Benusiglio and C. Clanet for very interesting
discussions.



\begin{thebibliography}{0}
\bibitem{overlooked}
Note that the particular case where the typical size $b$ is much smaller than the capillary length
$\kappa^{-1}=({\gamma/\rho g})^{1/2}$ has also been (briefly) discussed 
in \cite{Elie:96}.
It has been shown that the wave resistance displays a maximum
at a velocity of the order of $\sqrt{\gamma / (\rho b)}$ \cite{Elie:96}.
However, the consequences of a pressure distribution of finite size on the 
amplitude of the discontinuity of the wave resistance at $V = c_{\rm min}$ 
has been overlooked in \cite{Elie:96}.
\bibitem{RefASteinberg} Note that the effect of a pressure distribution of finite size 
on the amplitude of the discontinuity of the wave resistance was
already briefly discussed in \cite{Steinberg1}.
\bibitem{VmaxDisp} Note however, that $V_{{\rm max}}$ is not simply given by $\sqrt{\gamma/(\rho b) + gb}$
\bibitem{approx} For velocities up to $0.8 \sqrt{g b}$, on can show that equation (\ref{pure}) is well
approximated by $p^2\kappa(\pi \gamma^{-1})\sqrt{\pi/2}\exp(-1/x^4)/(2x^4)$ with $x=V/\sqrt{gb}$.
\bibitem{nonsym} In the context of animal locomotion on water surfaces, it would be interesting to generalize the results of the present study to non-axisymmetric pressure sources.
\end{thebibliography}
\end{document}